%Paper: gr-qc/9412010
%From: <moffat@medb.physics.utoronto.ca>
%Date: Fri, 2 Dec 1994 15:53:19 -0500

%This is a plainTeX macro to process manuscript called paper.tex
\magnification=1200
\voffset=0 true mm
\hoffset=0 true in
\hsize=6.5 true in
\vsize=8.5 true in
\normalbaselineskip=13pt
\def\doublespace{\baselineskip=20pt plus 3pt\message{double space}}
\def\singlespace{\baselineskip=13pt\message{single space}}
\let\spacing=\singlespace
\parindent=1.0 true cm

% bold face mathe italic fonts in dir 2160, 1800, 1643, and 1500
 %ambi in VAX

% also available in dir 1000,1095,1200,1315, and 1440

\newcount\equationumber \newcount\sectionumber
\sectionumber=1 \equationumber=1
\def\setsection{\global\advance\sectionumber by1 \equationumber=1}

\def\numbe{{{\number\sectionumber}{.}\number\equationumber}
                            \global\advance\equationumber by1}
\def\numberit{\eqno{(\number\equationumber)} \global\advance\equationumber by1}

\def\numberal{(\number\equationumber)\global\advance\equationumber by1}

\def\sectionit{\eqno{(\numbe)}}

\def\ccf#1{\,\vcenter{\normalbaselines
    \ialign{\hfil$##$\hfil&&$\>\hfil ##$\hfil\crcr
      \mathstrut\crcr\noalign{\kern-\baselineskip}
      #1\crcr\mathstrut\crcr\noalign{\kern-\baselineskip}}}\,}
\def\scf#1{\,\vcenter{\baselineskip=9pt
    \ialign{\hfil$##$\hfil&&$\>\hfil ##$\hfil\crcr
      \vphantom(\crcr\noalign{\kern-\baselineskip}
      #1\crcr\mathstrut\crcr\noalign{\kern-\baselineskip}}}\,}

\def\small3j#1#2#3#4#5#6{\def\st{\scriptstyle} % 3j-symbol - small size
   \bigl(\scf{\st#1&\st#2&\st#3\cr
           \st#4&\st#5&\st#6\cr} \bigr)}

   %Name of a nucleus

%\def\slashA{\hbox{$A\mkern-9mu/\mkern 9mu$}}

%\def\slasshA{\hbox{$A\mkern-9mu/\mkern 5mu$}}

\def\ref#1{$^{#1)}$}

   %Figure caption
              %#4 for caption

%...... subscripts and supscripts .....................................
\def\upa#1{\raise 1pt\hbox{\sevenrm #1}}
\def\dna#1{\lower 1pt\hbox{\sevenrm #1}}
\def\dnb#1{\lower 2pt\hbox{$\scriptstyle #1$}}
\def\dnc#1{\lower 3pt\hbox{$\scriptstyle #1$}}
\def\upb#1{\raise 2pt\hbox{$\scriptstyle #1$}}
\def\upc#1{\raise 3pt\hbox{$\scriptstyle #1$}}
\def\hprime{\raise 2pt\hbox{$\scriptstyle \prime$}}
\def\ccom{\,\raise2pt\hbox{,}}

%.... special maths symbols

\def\asymptotically#1{\;\rlap{\lower 4pt\hbox to 2.0 true cm{
    \hfil\sevenrm  #1 \hfil}}
   \hbox{$\relbar\joinrel\relbar\joinrel\relbar\joinrel
     \relbar\joinrel\relbar\joinrel\longrightarrow\;$}}
\def\Asymptotically#1{\;\rlap{\lower 4pt\hbox to 3.0 true cm{
    \hfil\sevenrm  #1 \hfil}}
   \hbox{$\relbar\joinrel\relbar\joinrel\relbar\joinrel\relbar\joinrel
     \relbar\joinrel\relbar\joinrel\relbar\joinrel\relbar\joinrel
     \relbar\joinrel\relbar\joinrel\longrightarrow$\;}}

\def\dal{\mathop{\sqcup\hskip-6.4pt\sqcap}\nolimits}

\catcode`@=11
\def\C@ncel#1#2{\ooalign{$\hfil#1\mkern2mu/\hfil$\crcr$#1#2$}}
\def\gf#1{\mathrel{\mathpalette\c@ncel#1}}      % slash a small letter
\def\Gf#1{\mathrel{\mathpalette\C@ncel#1}}      % slash a big letter

\def\gapx{\lower 2pt \hbox{$\buildrel>\over{\scriptstyle{\sim}}$}}
\def\lapx{\lower 2pt \hbox{$\buildrel<\over{\scriptstyle{\sim}}$}}

\def\nablaleft{\hbox{\raise 6pt\rlap{{\kern-1pt$\leftarrow$}}{$\nabla$}}}
\def\nablaright{\hbox{\raise 6pt\rlap{{\kern-1pt$\rightarrow$}}{$\nabla$}}}
\def\nablaboth{\hbox{\raise 6pt\rlap{{\kern-1pt$\leftrightarrow$}}{$\nabla$}}}

\def\boks#1#2{{\hsize=#1 true cm\parindent=0pt
  {\vbox{\hrule height1pt \hbox
    {\vrule width1pt \kern3pt\raise 3pt\vbox{\kern3pt{#2}}\kern3pt
    \vrule width1pt}\hrule height1pt}}}}

\def\heading{ }
\def\range{ }

\def\body{\vfill\eject\parindent=1.0 true cm
 \ifx\spacing\singlespace\singlespace\else\doublespace\fi}
\def\title#1{\centerline{{\bf #1}}}

\def\today{\ifcase\month\or
  January\or February\or March\or April\or May\or June\or
  July\or August\or September\or October\or November\or December\fi
  \space\number\day, \number\year}
\let\date=\today
\newcount\hour \newcount\minute
\countdef\hour=56
\countdef\minute=57
\hour=\time
  \divide\hour by 60
  \minute=\time
  \count58=\hour
  \multiply\count58 by 60
  \advance\minute by -\count58

\def\sectionskip{\penalty-500\vskip24pt plus12pt minus6pt}

\def\sec{\hbox{\lower 1pt\rlap{{\sixrm S}}{\hbox{\raise 1pt\hbox{\sixrm S}}}}}
\def\section#1\par{\goodbreak\message{#1}
    \sectionskip\nobreak\noindent{\bf #1}\vskip0.3cm \noindent}

\nopagenumbers
\headline={\ifnum\pageno=\count31\frontheadline
  \else{\ifnum\pageno=0\frontheadline
     \else{{\raise 2pt\hbox to \hsize{\paperhead}}}\fi}\fi}
%\headline={\ifnum\pageno=\count31\frontheadline
%  \else{\ifnum\pageno=0\frontheadline
%     \else{\underbar{\raise 2pt\hbox to \hsize{\paperhead}}}\fi}\fi}

\footline={\centerline{\sevenbf \folio}}
\def\frontheadline{\sevenbf \hfil}
\def\paperhead{\sevenbf \heading\ \range\hfil\folio}
\newdimen\pagewidth \newdimen\pageheight \newdimen\ruleht
\maxdepth=2.2pt
\pagewidth=\hsize \pageheight=\vsize \ruleht=.5pt

\def\onepageout#1{\shipout\vbox{ % here we define one page of output
    \offinterlineskip % butt the boxes together
  \makeheadline
    \vbox to \pageheight{
         #1 % now insert the main information
 \ifnum\pageno=\count31{\vskip 21pt\line{\the\footline}}\fi
 \ifvoid\footins\else %footnot ino is present
 \vskip\skip\footins \kern-3pt
 \hrule height\ruleht width\pagewidth \kern-\ruleht \kern3pt
 \unvbox\footins\fi
 \boxmaxdepth=\maxdepth}
 \advancepageno}}
\output{\onepageout{\pagecontents}}
\count31=-1
\topskip 0.7 true cm
%end of TeX macro
\pageno=0
\doublespace
\centerline{\bf How Old is the Universe?}
\vskip 1 true in
\centerline{\bf J. W. Moffat}
\centerline{\bf Department of Physics}
\centerline{\bf University of Toronto}
\centerline{\bf Toronto, Ontario M5S 1A7, Canada}
\vskip 0.5 true in
\centerline{\bf ABSTRACT}
\vskip 0.2 true in
A dynamical model of the decaying vacuum energy is presented, which is based
on Jordan-Brans-Dicke theory with a scalar field $\phi$. The solution of
an evolutionary differential equation for the scalar field $\phi$ drives the
vacuum energy towards a cosmological constant at the present epoch
that can give for the age of the universe, $t_0\sim 13.5$ Gyr for $\Omega_0=1$,
which is consistent with the age of globular clusters.
\vskip 1 true in
UTPT-94-34
\vskip 0.3 true in
e-mail: moffat@medb.physics.utoronto.ca
\par\vfil\eject
\proclaim 1. {\bf Introduction}\par
\vskip 0.2 true in
The age of the universe problem is a well-known quandary of
the spatially homogeneous models. Estimates of the age of the Galaxy based
on radioactive dating, the age of globular clusters and other methods give
the age of the universe within 10 to 20 Gyr$^{1}$.  At the same time, in
order for a matter dominated FRW model with the density parameter
$\Omega_0 = 1$ to yield $t_0 \geq 10$ Gyr requires $h \leq
0.65$, where $h$ is related to the observationally established value of the
Hubble constant through $H_0 =100 h \hbox{km} {\hbox{s}}^{-1}
{\hbox{Mpc}^{-1}}$. A value of $h \geq 0.65$ produces the age of the
universe problem. Recent measurements using the Canada-France-Hawaii
Telescope$^{2}$ and the Hubble Space Telescope$^{3}$ have given
$H_0=87\pm 7 \hbox{km} {\hbox{s}}^{-1}
{\hbox{Mpc}^{-1}}$ and $80\pm 17 \hbox{km} {\hbox{s}}^{-1}
{\hbox{Mpc}^{-1}}$, respectively.
These results imply an age for the universe of $7\times 10^9$ yrs and
$8\times 10^9$ yrs, respectively, in the standard big bang model with
$t_0=2/3\times 1/H_0$. These results have produced a new age of the universe
crisis.

The inflationary paradigm$^{4}$ has motivated the
need to have $\Omega_0=1$. Also, dynamical estimates of the FRW density
parameter ${\Omega}_{0}$ give results for observations over large scales
($>20 \hbox{ Mpc}$, say $\sim 100 \hbox{ Mpc }$) indicating
the existence of a less clustered component with a contribution
perhaps as high as ${\Omega}_{\sim 100} \simeq 0.8 \pm 0.2\,$$^{5}$.

The standard solution to the age of the universe problem
is based on the postulate that the cosmological constant $\Lambda$ is
non-zero$^{6}$. For $\Omega_{\hbox{matter}}=0.2$ and
$\Omega_{\hbox{vacuum}}=0.8$, corresponding to $\Omega
=\Omega_{\hbox{matter}}+\Omega_{\hbox{vacuum}}=1$, the expansion age is
$t_0=13.5$ Gyr, which is consistent with the other estimates of the age.
But the cosmological constant is very small, $\Lambda <
10^{-46}\,\hbox{GeV}^4$,
so its use to obtain a consistent age of the universe introduces a severe
fine-tuning problem.

The cosmological constant provides a source of negative pressure,
producing an acceleration of the expansion of the universe that negates
the deceleration of the expansion caused by matter and radiation. This
makes the universe appear younger than it is to an observer who assumes
that the comsological constant is zero. If $H_0t_0$ exceeds unity, it
means the expansion has been accelerating rather than decreasing, and within
the
standard model this would imply that the active gravitational mass density
is negative, which would happen if there were an appreciable positive
cosmological constant.

In Einstein's field equations:
$$
G_{\mu\nu}+\Lambda_B g_{\mu\nu}=8\pi GT_{\mu\nu},
\sectionit
$$
$\Lambda_B$ is called the ``bare'' cosmological constant, while a second
cosmological constant enters through the vacuum energy density:
$$
T_{V\mu\nu}=-\rho_Vg_{\mu\nu}=-{\Lambda_V\over 8\pi G}g_{\mu\nu}.
\sectionit
$$
These two constants can be combined through an effective cosmological constant:
$$
\Lambda=\Lambda_B+\Lambda_V.
\sectionit
$$
Why $\Lambda$ has the incredibly small value, $\Lambda < 10^{-46}\hbox{GeV}^4$,
today is known as the cosmological constant problem$^{7}$.

We shall seek an explanation for the smallness of $\Lambda$ at the present
epoch, and the fact that it has a critical value needed to resolve the
age of the universe problem by using the Jordan-Brans-Dicke (JBD)
action$^{8-12}$.
In JBD, the kinetic energy can produce a negative pressure, a fact emphasized
recently by Levin and Freese$^{13}$ in connection with inflationary theory. The
nonminimal coupling of the scalar
field to gravity allows for a negative pressure associated with the kinetic
energy of the field. By assuming that $\Lambda$ is spacetime dependent and
equating it to the kinetic energy in the JBD theory, we find that
the evolutionary equation for the scalar field in an FRW universe, produces an
attractor mechanism
that drives $\Lambda$ towards a minimum of the potential. This damping
mechanism can give a small constant value for $\Lambda$ today that can
resolve the age of the universe problem.

We shall assume that $\Lambda_B=0$ and
that the vacuum energy is dominated by the scalar field $\phi$. Unbroken
supersymmetry can ``protect'' the constant $\Lambda_B$ from becoming
non-zero$^{7}$,
but in the real universe supersymmetry will be broken at an energy $> 1$
Tev. We do not know of any other symmetry that guarantees that $\Lambda_B$
remains zero.
\vskip 0.2 true in
\setsection\proclaim 2. {\bf Tensor-Scalar Theory}\par
\vskip 0.2 true in
In the Jordan-Fierz conformal frame the Jordan-Brans-Dicke
action for a massless tensor-scalar theory takes the form:
$$
{\tilde S}={1\over 16\pi}\int d^4x\sqrt{-{\tilde g}}\biggl[\psi{\tilde R}
-{\omega(\psi)\over \psi}{\tilde g}^{\mu\nu}\psi_{,\mu}\psi_{,\nu}\biggr]
+S_m,
\sectionit
$$
where ${\tilde R}={\tilde g}^{\mu\nu}{\tilde R}_{\mu\nu}$ denotes the curvature
scalar in terms of the physical metric tensor ${\tilde g}_{\mu\nu}$,
$\omega(\psi)$ denotes the JBD $\psi$-dependent coupling parameter and
$S_m$ denotes the action for matter. In the ``Einstein frame'' the action
reads:
$$
S={1\over 16\pi G}\int d^4x\sqrt{-g}(R-2g^{\mu\nu}\phi_{,\mu}\phi_{,\nu})+S_m,
\sectionit
$$
where $g_{\mu\nu}$ denotes the ``Einstein'' metric tensor conformally
related to the physical Jordan-Fierz one by the transformation:
$$
{\tilde g}_{\mu\nu}=C^2(\phi)g_{\mu\nu}
\sectionit
$$
and $R=g^{\mu\nu}R_{\mu\nu}$.

Let us now define the scalar field:
$$
\Lambda(x)=g^{\mu\nu}\phi_{,\mu}\phi_{,\nu},
\sectionit
$$
and identify $\Lambda(x)$ as a variable cosmological ``constant''. Then,
Eq.(2.2) becomes:
$$
S={1\over 16\pi G}\int d^4x\sqrt{-g}(R-2\Lambda)+S_m.
\sectionit
$$
The field equations in the Einstein conformal frame are:
$$
R_{\mu\nu}=2\Lambda_{\mu\nu}+8\pi G(T_{\mu\nu}-{1\over 2}g_{\mu\nu}T),
\sectionit
$$
$$
\dal_g\phi=-4\pi GaV(\phi)T,
\sectionit
$$
where $T_{\mu\nu}$ denotes the Einstein stress-energy tensor,
$\dal_g=g^{\mu\nu}\nabla_\mu\nabla_\nu$ and $\nabla_\mu$ is the covariant
derivative with respect to the $g_{\mu\nu}$ connection. Moreover,
$$
\Lambda_{\mu\nu}=\phi_{,\mu}\phi_{,\nu}
\sectionit
$$
and
$$
a(\phi)\equiv {\partial V(\phi)\over \partial\phi}
\sectionit
$$
is the gradient of the coupling function $V(\phi)$ where
$$
V(\phi)=\hbox{ln}C(\phi).
\sectionit
$$
\vskip 0.2 true in
\setsection\proclaim 3. {\bf Tensor-Scalar Cosmology with Varying
$\Lambda$}\par
\vskip 0.2 true in
The metric takes its usual FRW form:
$$
ds^2=dt^2-R^2(t)\biggl[{dr^2\over 1-kr^2}+r^2(d\theta^2+\hbox{sin}^2\theta
d\phi^2)\biggr],
\sectionit
$$
where $k=+1,0,-1$ denotes the sign of the spatial curvature. The matter
distribution is described by a perfect fluid form:
$$
T^{\mu\nu}=(\rho+p)u^\mu u^\nu-pg^{\mu\nu}
\sectionit
$$
with $g_{\mu\nu}u^\mu u^\nu={\tilde g}_{\mu\nu}{\tilde u}^\mu {\tilde
u}^\nu=1$,
$\rho=C^4{\tilde \rho}$ and $p=C^4{\tilde p}$. The Jordan-Fierz
variables ${\tilde t}$ (physical cosmic time) and ${\tilde R}$ are given by
$$
d{\tilde t}=C(\phi(t))dt,
\sectionit
$$
$$
{\tilde R}({\tilde t})=C(\phi(t))R(t).
\sectionit
$$

The Einstein equations (2.6) give
$$
{{\ddot R}\over R}=-{4\pi\over 3}G(\rho+3p)-{2\Lambda\over 3},
\sectionit
$$
$$
\biggl({{\dot R}\over R}\biggr)^2+{k\over R^2}
={8\pi G\rho\over 3}+{\Lambda\over 3},
\sectionit
$$
where $\Lambda=({\dot\phi})^2$ and ${\dot\phi}=d\phi/dt$. The scalar
field equation is
$$
{\ddot\phi}+({\dot\phi})^2+3{{\dot R}\over R}{\dot\phi}
=-4\pi Ga(\phi)(\rho-3p).
\sectionit
$$
The conservation equation reads:
$$
d(\rho R^3)+pdR^3=(\rho-3p)R^3dV(\phi).
\sectionit
$$

We introduce the evolution parameter:
$$
\tau=\hbox{ln}R+\hbox{const},
\sectionit
$$
and $d\tau=Hdt$ where $H={\dot R}/R$ is the Hubble parameter in the Einstein
frame. We also have
$$
{\dot\phi}=H{d\phi\over d\tau}.
\sectionit
$$
\vskip 0.2 true in
\setsection\proclaim 4. {\bf Scalar Field Evolution Equation}\par
\vskip 0.2 true in
The evolution equation for the scalar field $\phi$ for $k=0$ takes the
form:
$$
{2\over 3-\phi'^2}\phi''+(1-\theta)\phi'=-(1-3\theta)a(\phi),
\sectionit
$$
where $\phi'=d\phi/d\tau$ and $\theta
={\tilde p}/{\tilde \rho}$ is considered to be either a constant or a known
function of $\tau$.

The trajectories in the $\phi$, ${\dot\phi}$ phase space associated with the
solutions of Eq.(4.1) have been studied extensively by Belinskii et al.,
$^{14}$ and by other authors$^{15-18}$. They showed that the point of origin,
$\phi,{\dot\phi}=(0,0)$,
is a point of stable focus and is the only point of equilibrium, while other
singular points can lie only at infinity.

Eq.(4.1) describes the motion of a ``$\phi$'' particle in a curved
spacetime with a limiting
speed equal to $\sqrt{3}$, a friction term proportional to the velocity,
$\phi'$, and a force term proportional to the gradient of the potential,
$V(\phi)$. For a dominant energy condition, the physical energy density
${\tilde\rho}$ is positive and $-1 < \theta < +1$, so that the second term in
(4.1) always damps the motion of the particle. For an inflationary era
the force term on the right-hand side of (4.1) is negative ($\theta =-1$);
it nearly vanishes for the
radiation dominated era ($\theta\approx 1/3$) and is negative for the matter
dominated one ($\theta\approx 0$).

In the radiation dominated era, Eq.(4.1) gives to first approximation:
$$
{2\over 3-\phi'^2}\phi''+{2\over 3}\phi'=0.
\sectionit
$$
A solution of (4.2) is$^{18}$
$$
\phi(\tau)=\phi_{\infty}-\sqrt{3}\hbox{ln}[B\hbox{exp}(-\tau)
+(1+B^2\hbox{exp}(-2\tau))^{1/2}],
\sectionit
$$
where
$$
B={\phi'_0/\sqrt{3}\over \sqrt{1-\phi_0'^2/3}}.
\sectionit
$$
The total displacement of a particle between $\tau=0$ and
$\tau=\infty$ is given by
$$
\Delta\phi=\phi_{\infty}-\phi_0
={1\over 2}\sqrt{3}\hbox{ln}{1+\phi'_0/\sqrt{3}\over 1-\phi'_0/\sqrt{3}}.
\sectionit
$$
For non-relativistic motion, $\phi'_0 << \sqrt{3}$, we get
$$
\Delta\phi\approx \phi'_0.
\sectionit
$$
Thus, upon entering the radiation era the $\phi$ particle will fall down the
potential $V(\phi)$ by a small amount.

For the matter dominated era with $\theta={\tilde p}/{\tilde \rho} << 1$,
the evolution equation becomes
$$
m(\phi')\phi''+\phi'=-a(\phi),
\sectionit
$$
with the mass:
$$
m(\phi')={2\over 3-\phi'^2}.
\sectionit
$$
Here $m(\phi')$ has the non-relativistic limit $m(\phi')\approx 2/3$. The
$\phi$ particle enters the matter era with an approximately zero initial
velocity and $\sqrt{3}$ is a critical value for $\vert\phi'\vert$. The
long-term behaviour of the particle depends dominantly on the shape of the
potential $V(\phi)$ and {\it the particle ends up near a local minimum of}
$V(\phi)$.

The behavor of the solution to Eq.(4.7) can be analyzed for the simple
case of a parabolic potential$^{18}$:
$$
V(\phi)={1\over 2}\kappa\phi^2,
\sectionit
$$
where $\kappa$ is the curvature:
$$
\kappa(\phi)={\partial a(\phi)\over \partial\phi}
={\partial^2V\over \partial\phi^2}
\sectionit
$$
calculated at the bottom of the well. For the potential (4.9), $\kappa(\phi)
=\kappa$ is a constant. The solution to Eq.(4.7) has a critical
value, $\kappa=3/8$, and for $0 < \kappa < 3/8$ the motion is monotonic and
is a linear combination of decreasing exponentials with positive coefficients.
For $\kappa > 3/8$, the motion is a damped oscillatory one and when
$\kappa\rightarrow 3/8$, the solution tends towards the critically damped
solution:
$$
\phi(\tau)=\phi_R(1+{3\over 4}\tau)\hbox{exp}(-{3\over 4}\tau),
\sectionit
$$
where $\phi_R$ is the value of $\phi$ at the end of the radiation era.

We expect that after a long time the particle is forced towards a minimum
of $V(\phi)$ (near where $a(\phi)$ vanishes). Thus, an {\it attractor
mechanism} forces the variable cosmological term $\Lambda$ towards small
or zero values.
\vskip 0.2 true in
\setsection\proclaim 5. {\bf Present Value of the Cosmological Constant}\par
\vskip 0.2 true in
In the matter dominated era, we can rewrite Eq.(3.6) in terms of the
Jordan-Fierz variables and $\phi(\tau)$ (recall that $\Lambda=({\dot\phi})^2$):
$$
{\tilde\Omega}_M={(1+a^2)(1-\phi'^2/3)\over (1+a\phi')^2}
+{(1+a^2)k\over {\tilde H}^2{\tilde R}^2},
\sectionit
$$
where
$$
{\tilde\Omega}_M={8\pi{\tilde G}{\tilde\rho}_M\over 3{\tilde H}^2}.
\sectionit
$$
Here, ${\tilde\rho}_M$ is the matter energy density and
$$
{\tilde G}(\phi)=GC^2(\phi)[1+a^2(\phi)].
\sectionit
$$

The observational limits on $a$ are (1$\sigma$ level):
$$
a^2_{\hbox{solar system}} < 0.01.
\sectionit
$$
Then (5.1) gives approximately for $k=0$ at the present epoch:
$$
{\tilde\Omega}_0={\tilde\Omega}_{0M}+{\tilde\Omega}_{0V}\approx 1,
\sectionit
$$
where from (3.3), (3.4) and (3.10), we have
$$
{\tilde\Omega}_{0V}={\phi'^2_0\over 3}
={(d\phi_0/d{\tilde t})^2\over 3{\tilde H}_0^2},
\sectionit
$$
which can be written as
$$
{\tilde\Omega}_{0V}={8\pi G{\tilde\rho}_{0V}\over 3{\tilde H}_0^2}
={{\tilde\Lambda}_0(\phi_0)\over 3{\tilde H}_0^2},
\sectionit
$$
where ${\tilde{\rho}}_{0V}$ denotes the vacuum energy density at the present
epoch.

We now have a scenario in which the scalar velocity is damped out
to its present value $\phi'_0$ at a minimum of the potential
$V(\phi)$. This corresponds to having an initially large cosmological
term ${\tilde\Lambda}(x)$
during the inflationary period in the early universe, which is damped out
in the matter dominated era to its present value ${\tilde\Lambda}_0$.

In the standard picture of inflation, the cosmic inflation
is driven by the potential for the scalar field $\phi$ (i.e.,
$\phi''\approx 0$ in Eq.(4.1)), whereas we have
the opposite situation in which negative pressure is produced by the kinetic
energy of the JBD theory.  We shall assume that the standard potential-driven
scenario will hold at near Planckian times, producing an inflationary
expansion. Then, as the universe enters the non-inflationary era, the
${\phi''}$ term in (4.1) becomes important and we have the scenario which
drives $\Lambda$ to
small or zero values as $\phi'$ approaches the minimum of $V(\phi)$.
\vskip 0.2 true in
\setsection\proclaim 6. {\bf The Age of the Universe}\par
\vskip 0.2 true in
For $k=0$ and a dust-dominated universe, the present age of an object
that formed at a redshift $z_c$ is$^{7}$:
$$
t_c(z_c)={2\over 3}\biggl(1+{\Omega_M\over \Omega_V}\biggr)^{1/2}
H_0^{-1}\biggl\{\hbox{sinh}^{-1}\biggl({\Omega_V\over \Omega_M}\biggr)^{1/2}
-\hbox{sinh}^{-1}\biggl[\biggl({\Omega_V\over\Omega_M}\biggr)^{1/2}
(1+z_c)^{-3/2}\biggr]\biggr\}.
\sectionit
$$
If we choose $z_c=4$ and $\Omega_V/\Omega_M=9$ ($\Omega_M=0.1$), then
this yields $t_0=1.1\times H_0^{-1}=13.5$ Gyr instead of $t_0=
2/3H_0^{-1}=8$ Gyr ($\Omega_V=0$), which is not in conflict with the age of the
globular clusters$^{19,20}$.
\vskip 0.2 true in
\setsection\proclaim 7. {\bf Bounds on ${\tilde\Omega}_{0V}$}\par
\vskip 0.2 true in
The maximum possible value of ${\tilde\Omega}_{0V}$ is constrained by various
bounds produced by the decay of the vacuum and nucleosynthesis$^{18,21,22}$.
The total energy:
$$
E=-\hbox{ln}\biggl(1-{\tilde\Omega}_V\biggr)+V(\phi),
\sectionit
$$
must decrease between the radiation era and the present epoch giving:
$$
1-{\tilde\Omega}_{0V} > \hbox{exp}(V_0-V_R)={Q_0\over Q_R},
\sectionit
$$
where $Q_R$ denotes the value of a quantity at the end of the radiation era.

{}From nucleosynthesis, we can obtain the inequality:
$$
1-{\tilde\Omega}_{0V} > {1\over (1+a^2)^{1/2}\xi_{\hbox{nucleo}}}
\approx {1\over \xi_{\hbox{nucleo}}},
\sectionit
$$
where $\xi_{\hbox{nucleo}}$ denotes the speed-up factor: $\xi_{\hbox{nucleo}}
={\tilde H_{NS}}/{\tilde H}_{\hbox{standard}}$, where ${\tilde H}_{NS}$ denotes
the value of ${\tilde H}$ during nucleosynthesis and
${\tilde H}_{\hbox{standard}}$ is the value predicted by the standard big-bang
model. Because for standard big-bang nucleosynthesis,
$\xi_{\hbox{nucleo}}\sim 1$, this implies that ${\tilde\Omega}_{0V}\approx 0$.

In order to obtain a
large enough value of ${\tilde\Omega}_{0V}$ to accomodate the age of the
universe, we assume that the coupling of the scalar field $\phi$
to baryons and anti-baryons is small enough to weaken the bounds on
${\tilde\Omega}_{0V}$ obtained from particle physics and nuclear physics.
This will guarantee that the decay of the vacuum and nucleosynthesis do not
prevent ${\tilde\Omega}_{0V}$ from reaching a value of order $0.8-0.9$,
which is needed to obtain a big enough age of the universe.

Why does ${\tilde\Omega}_V$ end up today with the value
${\tilde\Omega}_{0V}\sim
0.8-0.9$ needed to give a reasonable age for the universe in the standard
model? We can only appeal to the efficiency of the damping mechanism,
corresponding to a special potential $V(\phi)$, and to the anthropic
principle as answers to this question. It can be argued that given our
knowledge of the fundamental constants of nature, the present vacuum density
could not be much smaller than it is today to sustain the large-scale structure
of the universe$^{7,23}$.
\vskip 0.2 true in
{\bf Acknowledgements}
\vskip 0.1 true in
I thank M. Clayton, J. Levin, G. Starkman and D. Yu. Pogosyan for helpful
and stimulating discussions. This work was supported by the Natural
Sciences and Engineering Research Council of Canada.
\vskip 0.3 true in
\centerline{\bf References}
\vskip 0.2 true in
\item{1.}{For references, see: P. J. E. Peebles, Principles of Cosmology,
Princeton University Press, New Jersey 1993, p.103.}
\item{2.}{M. J. Pierce et al., Nature {\bf 371}, 385 (1994).}
\item{3.}{W. L. Freedman et al., Nature, {\bf 371}, 757 (1994).}
\item{4.}{A. Guth, Phys. Rev. D {\bf 23}, 347 (1981); A. Linde, Rep. Prog.
Phys. {\bf 47}, 925 (1984).}
\item{5.}{G. Efstathiou et al., Mon. Not. R. Astron. Soc. {\bf 247}, 10 (1990);
W. Saunders et al., Nature, {\bf 349}, 32 (1991).}
\item{6.}{For a review, see: S. M. Carroll, W. H. Press and E. L. Turner,
Annu. Rev. Astrophys. {\bf 30}, 499 (1992).}
\item{7.}{S. Weinberg, Rev. Mod. Phys. {\bf 61}, 1 (1989).}
\item{8.}{P. Jordan, Nature {\bf 164}, 637 (1949); Z. Phys. {\bf 157},
112 (1959).}
\item{9.}{M. Fierz, Helv. Phys. Acta {\bf 29}, 128 (1956).}
\item{10.}{C. Brans and R. H. Dicke, Phys. Rev. {\bf 124}, 925 (1961).}
\item{11.}{P. G. Bergmann, Int. J. Theor. Phys. {\bf 1}, 25 (1968).}
\item{12.}{R. V. Wagoner, Phys. Rev. D {\bf 1}, 3209 (1970).}
\item{13.}{J. J. Levin and K. Freese, Phys. Rev. D {\bf 47}, 4282 (1993);
Nucl. Phys. B {\bf 421}, 635 (1994); J. J. Levin, CITA preprints, 1994.}
\item{14.}{V. A. Belinskii et al., Sov. Phys. JETP {\bf 62}, 195 (1985); V. A.
Belinskii and I. M. Khalatnikov, Sov. Phys. JETP {\bf 66}, 441 (1987).}
\item{15.}{D. La and P. J. Steinhardt, Phys. Rev. Lett. {\bf 62}, 376
(1989).}
\item{16.}{J. D. Barrow and Kei-ichi Maeda, Nucl. Phys. B {\bf 341}, 294
(1990).}
\item{17.}{S. Kalara, N. Kaloper and K. A. Olive, Nucl. Phys. B {\bf 341},
252 (1990).}
\item{18.}{T. Damour and K. Nordtvedt, Phys. Rev. D {\bf 48}, 3436 (1993).}
\item{19.}{P. Demarque, C. P. Deliyannis and A. Sarajedini, Observational
Tests of Cosmological Inflation, ed. T. Shanks et al., 111. Dordrecht:
Kluwer (1991).}
\item{20.}{B. Chaboyer, CITA preprint, 1994.}
\item{21.}{M. \"Oser and M. O. Taha, Phys. Lett. {\bf 171}, 363 (1986);
K. Freese et al., Nucl. Phys. B {\bf 287}, 797 (1987).}
\item{22.}{B. Ratra and P. J. E. Peebles, Phys. Rev. D {\bf 37}, 3406 (1988).}
\item{23.}{G. Starkman, in preparation.}

\end